\documentclass{jaa}

\usepackage{graphicx}
\usepackage{hyperref}

\begin{document}\sloppy

\title{Study of dynamical status of the globular cluster NGC 1851\\ using Ultraviolet Imaging Telescope}


\author{Gaurav Singh\textsuperscript{1,2*}, R. K. S. Yadav\textsuperscript{1}, Snehalata Sahu\textsuperscript{3} and Annapurni Subramaniam\textsuperscript{3}}
\affilOne{\textsuperscript{1}Aryabhatta Research Institute of Observational Sciences (ARIES), Manora Peak, Nainital, 263001\\}
\affilTwo{\textsuperscript{2}Department of Physics and Astrophysics, University of Delhi, Delhi 110007\\}
\affilThree{\textsuperscript{3}Indian Institute of Astrophysics, Koramangala II Block, Bangalore-560034, India}

\twocolumn[{

\maketitle

\corres{gaurav@aries.res.in}


\begin{abstract}
We present the study of dynamical status of the globular cluster NGC 1851. A combination of multi-wavelength space and ground-based data sets are used for the present analysis. In order to select the genuine cluster members, we used the astro-photometric data available from HST and GAIA-DR2 catalogs. The BSS radial distribution of the cluster is plotted from the center of the cluster to the outskirts. The radial distribution of BSS shows a central peak, followed by a dip at the intermediate radii ($r_{min}$ $\sim$ $90^{''}$) and a rising trend in the outskirts. We also estimated $A^{+}_{rh}$ parameter as 0.391 $\pm$ 0.006 to validate the findings of the radial distribution study. On the basis of the minima in the BSS radial distribution and the value of $A^{+}_{rh}$ parameter, we conclude that NGC 1851 belongs to \textit{Family II} classification and is an intermediate dynamical state cluster.
\end{abstract}


}]


\doinum{12.3456/s78910-011-012-3}
\artcitid{\#\#\#\#}
\volnum{000}
\year{0000}
\pgrange{1--}
\setcounter{page}{1}
\lp{1}

\section{Introduction}

Globular clusters (GCs) are compact, centrally concentrated and gravitationally bound systems of stars. The high density in the central region of globular clusters lead to frequent gravitational interactions. The gravitational interactions among stars result in various dynamical processes i.e., two-body relaxation, core collapse, mass segregation,  stellar collisions, stellar mergers  etc (Meylan \& Heggie 1997). These dynamical processes give rise to several exotic populations i.e., millisecond pulsars, cataclysmic variables, and blue straggler stars (BSSs), etc (Ferraro {\em et al.} 2001; Bailyn 1995). BSSs are bright and are more populous among these exotic population, therefore serve as a crucial probe to understand the internal dynamics of the GCs. 

In 1953, Sandage discovered BSSs populating an unusual location in the Optical color-magnitude diagram (CMD) of the GC, M3 (Sandage 1953). He found the location of these stars to be bluer with respect to the main sequence stars and they appear as an extension of the  main sequence turnoff (MS-TO). Based on isochrone fitting technique, Shara {\em et al.} (1997) found BSSs to be more massive ($M$ $\sim$ 1.2 $M_{\odot}$) than the average mass of stars in the GCs ($M$ $\sim$ 0.3 $M_{\odot}$). Using Spectral energy distribution (SED) fitting technique, Raso {\em et al.} (2019) found similar estimates of BSS masses, with a few BSSs having masses larger that 2 times of MS-TO. Being more massive than the normal stars, they are subject to dynamical friction, which segregates the BSSs towards the centre of the cluster (Ferraro {\em et al.} 2006).

At the very centrally dense regions of GCs, HST astro-photometric catalog provides useful information to select cluster members. Also, after the release of GAIA $DR2$ it has now become feasible to select genuine BSS population from the outer region of the cluster. Once, the selection of genuine BSS population are completed, their radial distribution can be studied to infer the dynamical status of the GCs. On the basis of the shape of the observed radial distribution, Ferraro {\em et al.} (2012) (hereafter F12) classified GCs into three main categories;

\begin{enumerate}

\item \textit{Family I}: The radial distribution of BSSs show a flat distribution, and are classified as dynamically young systems. The examples of GCs classified in
\textit{Family I} are e.g., Palomar 14 (Beccari {\em et al.} 2011) , NGC 2419 (Dalessandro {\em et al.} 2008b), and $\omega$ Centauri (Ferraro {\em et al.} 2006). 

\item \textit{Family II}: The radial distribution of BSSs show a bimodal distribution with a central peak, followed by a minima at some intermediate radii ($r_{min}$), and an external rising trend. These clusters are classified as dynamically intermediate clusters. The example of the clusters showing bimodal distribution are: NGC 6388 studied by Dalessandro {\em et al.} (2008a), M53 by Beccari {\em et al.} (2008), M5 by Lanzoni {\em et al.} (2007a), and 47 Tuc by Ferraro {\em et al.} (2004), M55 by Lanzoni {\em et al.} (2007c), NGC 6752 by Sabbi {\em et al.} (2004) and NGC 5824 by Sanna {\em et al.} (2014).  

\item \textit{Family III}: The radial distribution of BSSs show a monotonic behaviour, with a central peak followed by a decreasing trend and no signs of an external rise. These clusters are classified as dynamically old systems. The examples of GCs showing unimodal behavior are e.g., M79 studied by Lanzoni {\em et al.} (2007b), M80 $\&$ M30 by F12, and M75 by Contreras Ramos {\em et al.} (2012).

\end{enumerate}

In the recent years, a new parameter ($A^{+}_{rh}$) has been proposed to measure the dynamical segregation of BSSs by Alessandrini {\em et al.} (2016), which is given as the area between the cumulative distribution curves of the reference population ($\phi_{REF}$(x)) and  BSSs ($\phi_{BSS}$(x)):

\begin{equation}
 A^{+}_{rh}(x) = \int^{x}_{x_{min}} \phi_{BSS}(x') - \phi_{REF}(x') dx'
\end{equation}

In the above equation, x (= log($r/r_{h}$)) is defined as the logarithmic distance from the center of the cluster and scaled over the half-mass radius $r_{h}$ of the cluster. Lanzoni {\em et al.} (2016) (hereafter L16) found a direct correlation between $r_{min}$ and $A^{+}_{rh}$ parameter, suggesting that both the parameters are governed by a basic mechanism i.e., dynamical friction.

UVIT/AstroSat observations have also been beneficial in the BSS radial distribution studies. Sahu {\em et al.} (2019) studied the specific frequency of BSS in the cluster NGC 288 and found a bimodal radial distribution using UVIT/AstroSat data.

In this paper, we present the study of dynamical status of the cluster, NGC 1851. This is a high density globular cluster with the core ($r_{c}$) and tidal ($r_{t}$) radii around $5^{''}.4$ and $481^{''}.6$, respectively (Ferraro {\em et al.} 2018). NGC 1851 ($\alpha_{J2000}$ = $5^{h}$ $14^{m}$ $6^{s}.76$, $\delta_{J2000}$ = $-40^{\circ}$ $2^{'}$ $47^{''}.6$; $l$ = $244^{\circ}$.51, $b$ = $-35^{\circ}$.03, Harris 1996 (2010 edition)  \footnote{\url{http://physwww.physics.mcmaster.ca/~harris/mwGC.dat}} (hereafter HA10)) is located at a distance of 12.1 kpc from the Sun. NGC 1851 is an intermediate metallicity cluster ([Fe/H] = $-$1.18). In Table 1, we list all the known properties of the cluster NGC 1851. Subramaniam {\em et al.} (2017) and Singh {\em et al.} (2020) presented the UV and Optical CMDs of the cluster using UVIT/AstroSat data.

The paper is organized in the following manner: data used for the present analysis is presented in Section 2, the selection of BSS and reference population followed by the results obtained from radial distribution in Section 3 and we discuss the results, followed by summary and conclusions in Sections 4 and 5.

\begin{table}
	\centering
	\caption{Parameters of the cluster NGC 1851 used in this paper.}
	\label{tab:decimal}
	\begin{tabular}{ccc}
		\hline
		\hline
		Parameter & Value & References\\
		\hline
		RA (J2000) & $5^{h}$ $14^{m}$ $6^{s}.76$ & HA10\\
		DEC (J2000) & $-40^{\circ}$ $2^{'}$ $47^{''}.6$ & HA10\\
		Distance & 12.1 kpc & HA10\\
		Metallicity, [Fe/H] & $-$ 1.18 dex & Kunder {\em et al.} (2013)\\
		Distance modulus & 15.47 mag & Cassisi {\em et al.} (2008)\\
		Age & 10 Gyr & Cassisi {\em et al.} (2008)\\
		Core radius ($r_{c}$) & $5^{''}.4$ & Ferraro {\em et al.} 2018\\
		Tidal radius ($r_{t}$) & $481^{''}.6$ & Ferraro {\em et al.} 2018\\
		\hline
		\hline
	\end{tabular}
\end{table}

\section{Data sets} \label{sec:style}
To study the dynamical status and to create the radial distribution of BSS in the cluster NGC 1851, we use space and ground-based data sets in NUV and Optical wavelengths. We aim to study the BSSs located in the entire cluster extension i.e., from the cluster centre to the tidal radii ($r_{t}$).

We use the astro-photometric catalog from Nardiello {\em et al.} (2018) in the central dense region ($r$ $\leq$ $90^{''}$) of the cluster, observed as a part of the HST UV Legacy Survey of Galactic Globular Cluster (HUGS) program (Piotto {\em et al.} 2015). The catalog contains the photometric data sets in \textit{F275W}, \textit{F336W}, and \textit{F438W} pass-bands, which were observed through WFC3/UVIS channel. The \textit{F606W} and \textit{F814W} pass-bands were observed through ACS/WFC channel. The catalog also contains the membership information of all the stars that are common to WFC3/UVIS and ACS/WFC field of view (FOV). 

To select the BSSs from the outer region ($90^{''}$ $\leq$ $r$ $\leq$ $480^{''}$), we use Near-UV (NUV) data in $\textit{N279N}$ filter, observed through {\it Ultra Violet Imaging Telescope (UVIT)} on board {\it AstroSat} satellite during $19^{th}$-$21^{st}$ March 2016, as a part of Performance Verification (PV) phase. UVIT/AstroSat provides a simultaneous observations in a wide range of electromagnetic spectrum from Far-UV (130-180 nm) to NUV (200-300 nm) wavelengths. It provides a circular field of view of about $\sim$ $28^{'}$ and an angular resolution better than $1.^{''}8$ in both the channels. Subramaniam {\em et al.} (2016) and Tandon {\em et al.} (2017) presented the details of instrument and calibration of the UVIT data. For this cluster, the data acquisition and reduction procedures are described in detail in Subramaniam {\em et al.} (2017). 

\begin{figure}[!t]
\includegraphics[width=1\columnwidth]{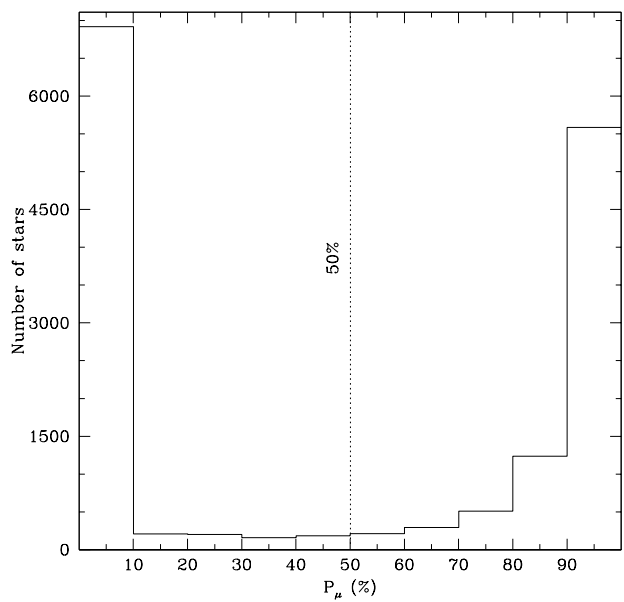}
\caption{The frequency vs membership probability for all the stars lying in the outer region is shown. In this histogram star counts start rising after 50 $\%$ membership probability. Therefore, we can use membership criteria of P $\geq$ 50 $\%$ for the selection of cluster members in the further analysis.}\label{figOne}
\end{figure}

For the optical wavelength in the outer region, we used the $\textit{UBVRI}$ photometric catalog provided by Stetson {\em et al.} (2019). The Gaia DR2 catalog provides the information of photometry and astrometry of all the stars down to G $\sim$ 21 mag (Gaia Collaboration {\em et al.} 2016, 2018). We, therefore, obtained the membership information using the method given in Sanders (1971) based on the maximum likelihood principle for all the stars located outside the HST region and using Gaia DR2 catalog. Figure \ref{figOne} shows the histogram plot (frequency vs membership probability) for all the stars lying in the outer region. Since, in the histogram star counts start rising after 50 $\%$ membership probability. We have used membership criteria of P $\geq$ 50 $\%$ for the selection of cluster members in the further analysis. So, by combining the HST membership catalog with the Gaia DR2 membership data, we now have membership information of most of the stars used in the present analysis.

\section{Analysis and Results } \label{sec:floats}

\subsection{BSS and reference population selection} \label{subsec:def}
The first step towards studying the BS radial distribution is to carefully select the genuine BSS and a reference population. For this purpose, we will use NUV pass-bands for the primary selection of the BSSs. To select the reference population and complete BSS population, we will use the Optical pass-bands for the selection, as described in (Singh \& Yadav 2019).

\subsection{BSS population selection}
For selecting the BSSs in the inner region, we use HST CMD (F275W, (F275W$-$F606W)) and for selecting BSSs from the outer region we use UVIT CMD (N279N, (N275N$-$V)). The NUV-Optical CMDs are fitted with the hybrid models i.e., the isochrone is fitted with the the Flexible Stellar Population Synthesis (FSPS) models of Conroy {\em et al.} (2009) and HB sequence is fitted with the HB model generated using the updated Bag of Stellar Tracks and Isochrones (BaSTI-IAC\footnote{\url{http://basti-iac.oa-abruzzo.inaf.it/hbmodels.html}}, Hidalgo {\em et al.} 2018). Both the FSPS and BaSTI-IAC model are generated for a metallicity of [Fe/H] = $-1.2$ dex (Kunder {\em et al.} 2013), a distance modulus of 15.47 mag and an assumed age of 10 Gyr (Cassisi {\em et al.} 2008). The theoretical isochrones are fitted well with the observed data points.

\begin{figure}[!t]
\includegraphics[width=1\columnwidth]{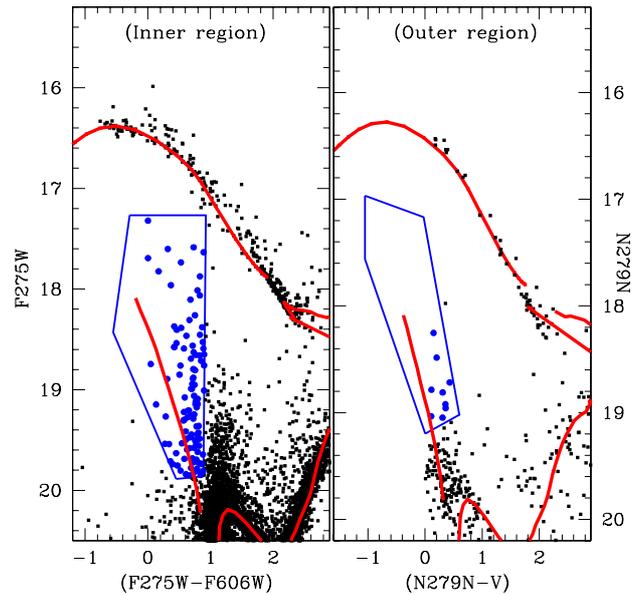}
\caption{The selection of BSSs from the NUV-Optical CMDs, with (F275W, (F275W$-$F606W)) in the $left$ $panel$ and (N279N, (N275N$-$V)) in the $right$ $panel$, respectively. The selected BSSs are shown with filled circles and they extends a vertical sequence in the NUV-Optical CMDs. The NUV-Optical CMDs are fitted with the hybrid models i.e., the isochrone is fitted with the FSPS models and the HB sequence is fitted with the HB model generated through BaSTI-IAC models. The selection box criteria is shown in both the panels.}\label{figTwo}
\end{figure}

Raso {\em et al.} (2017) have used UV CMD (F275W, (F275W$-$F336W)) to select the BSSs from the central region of the cluster. In the UV CMD, BSS sample can be selected both efficiently and reliably, since it can be easily separated from the optical blends just above the MS-TO. Therefore, to select the BSSs from the entire region, we use NUV-Optical CMD (HST and UVIT CMDs) as our primary selection criteria where the contamination from the optical blends can be minimized. In the NUV-Optical CMD, the BSSs can be separated from the optical blends which extends a vertical sequence just above the main sequence while in Optical CMD it is difficult to separate these two sequences. The selection box criteria is shown in the Figure \ref{figTwo}. Since, in NUV-Optical CMDs the BSSs can be easily distinguished and are separated from the optical blends and visible plumes. The evolutionary track of BSS in the theoretical isochones also provides useful information for defining the selection criteria for the selection of BSSs. In Figure \ref{figTwo}, the BSSs defines a vertical sequence, while the cooler giants like SGB, RGB are suppressed. The selected BSSs are shown with filled circles. To minimize the contamination from the MS-TO and the sub-giant branch, we adopted a limiting magnitude of F275W = 19.85 mag and N279N = 19.30 mag in the HST and UVIT CMDs, respectively. The upper limit extends up to the brightest BSSs found in both the samples. We have identified 154 BSSs in the inner region and 10 BSSs in the outer region using NUV-Optical CMDs. 

Once the BSSs are selected from the NUV-Optical CMDs, they are used to define the selection box criteria in the Optical CMD (V, (V$-$I)), as shown in Figure \ref{figThree}. By combining the Stetson's catalog with the \textit{Gaia} DR2 data and HST photometry, it is possible to cover both the inner and outer regions (Stetson {\em et al.} 2019). Therefore, we have used Stetson's catalog in the outer region, and HST in the inner region. In order to select the BSS and reference population from a common Optical CMD, the F606W and F814W magnitudes are transformed to calibrated V and I magnitudes available in the Stetson’s catalog, using the relationship derived by Sirianni {\em et al.} (2005). We found 3 additional BSSs from the Optical CMD in the inner region and 5 additional BSSs in the outer region. The reason for getting these additional BSSs is attributed to the fact that there are 3 BSSs that lies in the ACS FOV but are not present in the WFC3 FOV. However, the 4 additional BSSs in the outer region are due to the incompleteness of the UVIT N279N and one is the BSS+EHB photometric binary companion identified by Singh {\em et al.} (2020), that is located in the EHB sequence in the UVIT CMD.

Therefore, in total we have found 172 BSSs throughout the entire cluster region, 157 in the inner and 15 in the outer region. All the BSSs are genuine cluster members with a membership probability, $P$ $\geq$ 80 $\%$. 

\subsection{Reference population selection}
 Reference populations are important to understand the segregation of BSSs and we used post-MS stars, since they define a natural trend in radial distribution against which the radial distribution of the BSSs can be compared. The number of stars in a Post-MS stage is directly proportional to its  evolutionary time scales (Lanzoni {\em et al.} (2007c); Renzini \& Fusi Pecci (1988)). These Post-MS stars are, therefore, important for qualitative study of BSS specific frequency and radial distribution studies. The specific frequency of these branches are constant throughout the entire cluster region and are equal to the evolutionary time scales of the HB and GB (SGB+RGB) phase respectively. Therefore, these branches are considered as the reference population. 

\begin{figure}[!t]
\includegraphics[width=1\columnwidth]{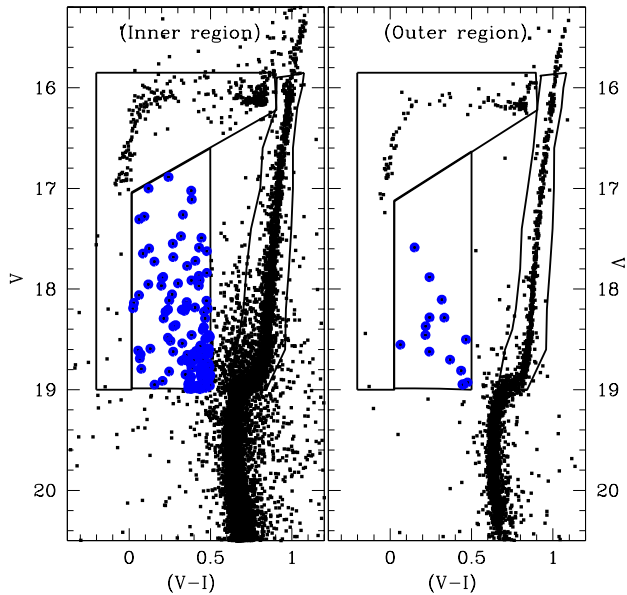}
\caption{The selection of the BSS and the reference population using Optical CMD (V, (V$-$I)). The selection box criteria of BSS and reference population are shown for both the inner and outer regions in the $left$ and the $right$ $panels$, respectively. The selected BSSs are marked by filled circles.}\label{figThree}
\end{figure}

For the reference population selection, we consider the same limiting magnitudes subtended by the BSS population in the optical CMD. Therefore, our selection of reference population are not affected by the completeness of the sample. To select the genuine reference population, we have adopted the membership criteria of $P$ $\geq$ 50 $\%$, since at higher membership cutoffs, we found a significant change in the number of reference population. In Figure \ref{figThree}, the selection criteria for reference population is shown. We used the same selection criteria followed by Singh \& Yadav (2019) to select the reference population. Hence, we found 394 and 127 HB stars in the inner and outer sample respectively. Also, 2690 and 634 GB stars from the inner and outer regions, respectively.

In total, we found, 521 HB and 3324 GB stars from the entire cluster region. 

\begin{figure}[!t]
\includegraphics[width=1\columnwidth]{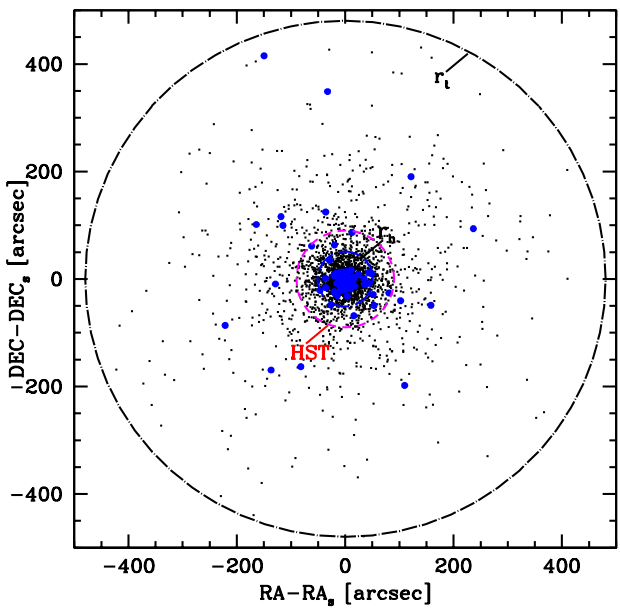}
\caption{The plot show the spatial distribution of BSSs in the entire cluster region and are marked by filled blue circles. The GB population is shown as black dots.}\label{figFour}
\end{figure}

Figure \ref{figFour}, show the spatial distribution of BSSs selected from the entire cluster region. The BSSs are more concentrated towards the center and a large fraction of them are located in the inner region, much within the half-mass radius ($r_{h}$). However, significant number of BSSs are present in the outer region of the cluster as well. We also plotted the spatial distribution of GB population which is nearly symmetric in comparison to the spatial distribution of BSSs.

\subsection{BSS radial distribution} \label{subsec:bss}
\begin{figure}[!t]
\includegraphics[width=1\columnwidth]{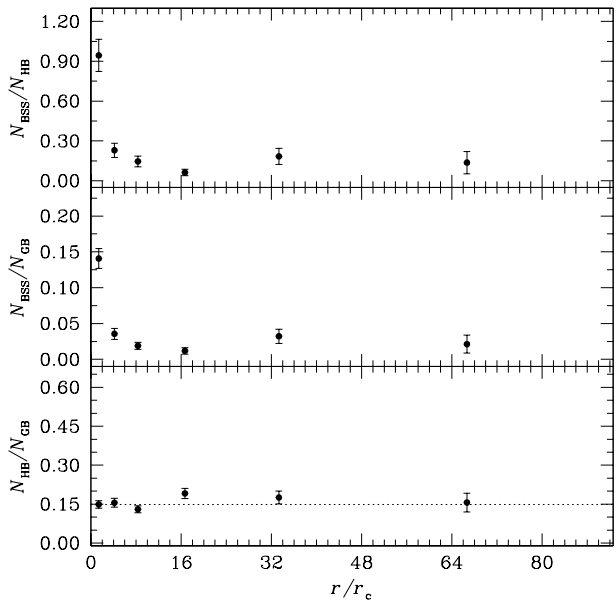}
\caption{The specific frequency of BSS with respect to HB and GB are plotted in the $upper$ and $middle$ $panels$, respectively, while the specific frequency of HB and GB is plotted in the $lower$ $panel$. The specific frequency distribution of BSS show a bimodal distribution, whereas for the reference population, it shows a constant value throughout the entire cluster region.}\label{figFive}
\end{figure}

In this section, we present the radial distribution of BSS with respect to the reference population. To obtain the radial distribution plot of BSS with respect to the reference population, we divided the cluster area into six concentric circles. In the Table 2, we listed the number of genuine BSS and reference population corresponding to each of the radial bin ($N_{BSS}$, $N_{GB}$ and $N_{HB}$.). We also obtain the specific frequency ($F^{BSS}_{GB}$ = $N_{BSS}$/$N_{GB}$, $F^{BSS}_{HB}$ = $N_{BSS}$/$N_{HB}$ and $F^{HB}_{GB}$ = $N_{HB}$/$N_{GB}$) of BSS and plotted them in Figure \ref{figFive}. The errors are plotted with 1 sigma error bar. The specific frequency of BSS show a bimodal distribution with a central peak followed by a minima and an external rising trend in the outer region, while the reference population show a flat radial distribution. In order to check the significance of the rise in the external region, we performed Z-test \footnote{\url{http://www.stat.yale.edu/Courses/1997-98/101/sigtest.htm}} and found the significance level of rise in the outer region to be $\sim$ 77 $\%$.

\begin{table}
	\centering
	\caption{The log of the Number counts for BSS and reference population.} 
	\label{tab:decimal}
	\begin{tabular}{cccccc}
		\hline
		\hline
		Radial bin &$N_{BSS}$ & $N_{HB}$ & $N_{GB}$ & $L_{samp}/L^{tot}_{samp}$\\
		(arcsec) & & & & \\ 
		\hline
		0 - 15 & 118 & 125 & 839 & 0.34\\
		15 - 30 & 18 & 96 & 619 & 0.20\\
		30 - 60 & 15 & 103 & 791 & 0.19\\
		60 - 120 & 7 & 113 & 591 & 0.15\\
		120 - 240 & 11 & 60 & 342 & 0.10\\
		240 - 480 & 3 & 22 & 141 & 0.06\\
		\hline
		\hline
	\end{tabular}
\end{table}

We also obtained the doubly normalized ratio for BSS with respect to the reference population. These population (``Pop'') could be BSS, HB or RGB. It is defined as the number of ``Pop'' observed in a region to the total number of ``Pop'' divided by the fraction of light sampled in the same region with respect to the total measured luminosity (Ferraro {\em et al.} 1993). It is written as,

\begin{equation}
R_{Pop} = \frac{N_{Pop}/N^{tot}_{Pop}}{L_{samp}/L^{tot}_{samp}}
\end{equation}

We estimated the sampled to the total luminosity for each radial bin by integrating the isotropic single-mass King profile using parameters taken from HA10 catalog. We assumed the Poisson error in the values of luminosities and numbers. Using the formula described in Sabbi {\em et al.} (2004), we considered propagation of errors to estimate the errors in the double normalized ratios. In Table 2, we listed the luminosity ratios computed in the corresponding annulus.

\begin{figure}[!t]
\includegraphics[width=1\columnwidth]{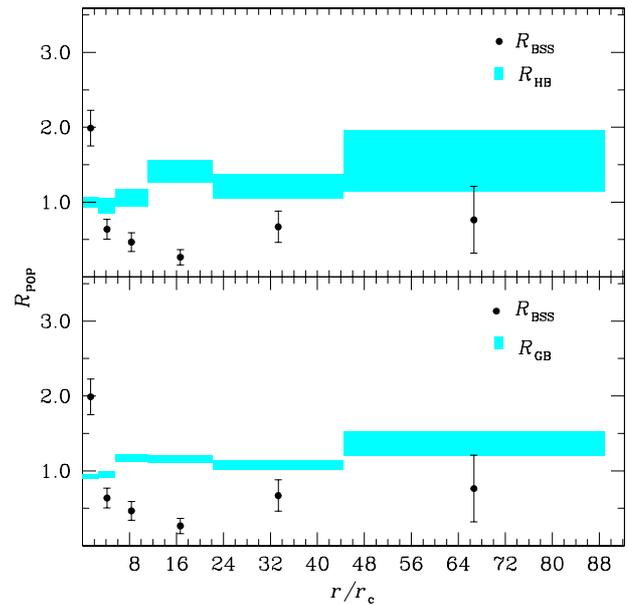}
\caption{The double normalized ratios of BSS are plotted with respect to HB and GB in the $upper$ and $lower$ $panels$, respectively. The BSS radial distribution show a peak in the center, a minima at the intermediate radii ($r_{min}$ $\sim$ $90^{''}$), and the external rising trend.}\label{figSix}
\end{figure}

In Figure \ref{figSix}, we plot the radial distribution of BSS, HB and GB, using double normalized ratios with respect to the radial distance scaled over $r_{c}$. In the $upper$ $panel$, double normalized ratio of BSS with respect to HB is plotted and the $lower$ $panel$ shows the variation of double normalized ratio of BSS with respect to GB. The value of $R_{BSS}$ shows a bimodal distribution with a peak in the center, a minima at $r$ $\sim$ 17 $r_c$ and an outward rising trend. The double normalized ratios of the reference population ($R_{HB}$ and $R_{GB}$), however show a flattened behaviour ($\sim$ 1), as expected from the stellar evolution theories (Renzini \& Fusi Pecci 1988).

\begin{figure}[!t]
\includegraphics[width=1\columnwidth]{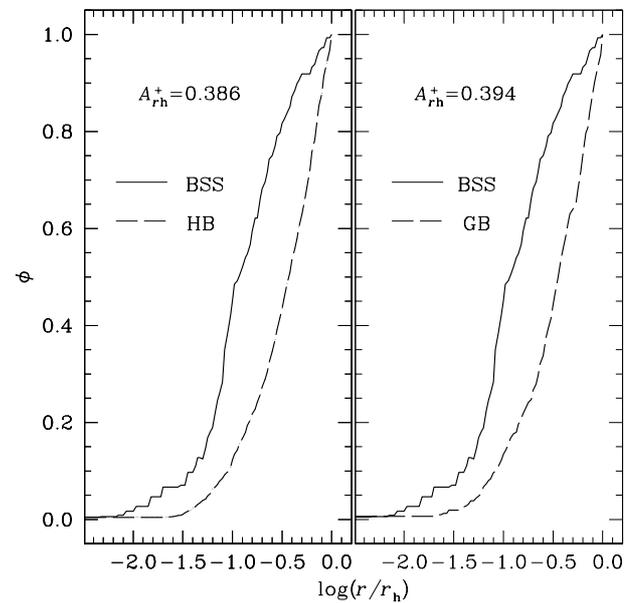}
\caption{The cumulative distribution plot of BSS, HB and GB. The $A^{+}_{rh}$ parameter is obtained as area between the curves of BSS with respect to HB and GB and are found to be 0.386 and 0.394, respectively. }\label{figSeven}
\end{figure}

\begin{figure}[!t]
\includegraphics[width=1\columnwidth]{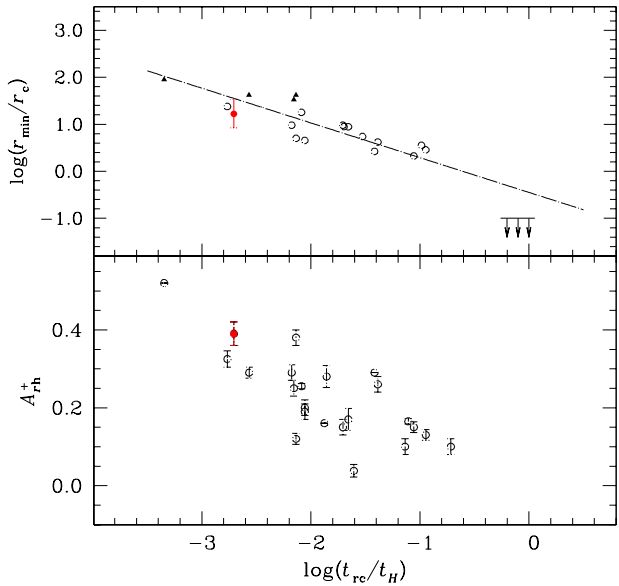}
\caption{The plot showing the location of the cluster NGC 1851 in the \textit{empirical dynamical clock} relation defined by F12 in the $upper$ $panel$. The filled triangles are dynamically old clusters while open circles are the dynamically intermediate age clusters. The dynamically young clusters are plotted as lower-limit arrows at $r_{min}$ $\sim$ 0.1. The cluster location in the empirical relation defined by L16 is shown in $lower$ $panel$.}\label{figEight}
\end{figure}

To recheck the dynamical segregation of the BSSs, we obtained the $A^{+}_{rh}$ parameter defined by L16, as the area enclosed between the cumulative radial distribution of BSS and the reference populations scaled over $r_{h}$. In Figure \ref{figSeven}, the cumulative radial distributions of BSS with respect to HB and GB are plotted in the $left$ and $right$ $panels$, respectively. The value of $A^{+}_{rh}$ are estimated as 0.394 and 0.386 with respect to GB and HB respectively. The corresponding values of mean and standard deviation of $A^{+}_{rh}$ are therefore $0.391$ and $0.006$, respectively.

In the $upper$ $panel$ of Figure \ref{figEight}, we plot the \textit{empirical dynamical clock} relation defined by F12, which correlates the position of minima of the BSS radial distribution ($r_{min}/r_{c}$) with the core relaxation time ($t_{rc}$/$t_{H}$). In the $lower$ $panel$ of Figure \ref{figEight}, we plot the correlation of $A^{+}_{rh}$ parameter and $t_{rc}$/$t_{H}$. The cluster NGC 1851 is shown with filled circle. The value of error in $r_{c}$ is taken from Miocchi {\em et al.} (2013) for estimating the error in the $r_{min}/r_{c}$. The values of $r_{min}$, $t_{rc}$, and $r_{c}$ of other clusters are adopted from F12. 

\section{Discussions}
The specific frequency and double normalized ratio of BSS show bimodal distribution, with a peak in the central region, followed by a clear dip at the intermediate radii ($r_{min}$ $\sim$ $90^{''}$) and a rising trend in the external region. The BSSs are more massive than the reference population and therefore are subjected to the dynamical friction which plays a crucial role in segregating the BSSs towards the cluster center. However, the BSSs that are located in the outskirts, show a flat distribution and are not yet affected by the action of dynamical friction. The radial distribution of BSS observed in NGC 1851 are in agreement with the previous studies (i.e., F12) on the cluster showing bimodal radial distribution. 

Ferraro {\em et al.} (2018) found the value of $A^{+}_{rh}$ parameter of the cluster NGC 1851 to be 0.48 $\pm$ 0.04. Our estimation of $A^{+}_{rh}$ parameter is slightly less than the value estimated by Ferraro {\em et al.} (2018), and it could be due to the strict condition of membership probability, $P$ $\geq$ 50 $\%$ for the selection of BSS and reference population. Although, Ferraro {\em et al.} (2018) included proper motion selection criteria using the Vector-Point diagram (VPD) from the HUGS data set. The difference in the $A^{+}_{rh}$ parameter could be largely due to the selection of reference population. Ferraro {\em et al.} (2018) used 5$\sigma$ selection box criteria for the selection of MS-TO as the reference population in the UV-CMD, while in the present analysis, we have done the selection of reference population (HB and GB) from the Optical CMD, due to incompleteness of the MS-TO sample in the UVIT CMD.

The $A^{+}_{rh}$ parameter of the NGC 1851, studied by Ferraro {\em et al.} (2018) does provides the information about the level of segregation of BSSs in the cluster sample, but in the central dense region of the cluster (up to $r_{h}$), therefore, BSS radial distribution is need to understand the level of segregation of BSSs located in the outskirts (after $r_{h}$) of the cluster. Also, it provides a very useful information about the cluster dynamical state. Therefore, we have estimated both the $r_{min}$ and $A^{+}_{rh}$ parameters to obtain the dynamical state of the cluster, NGC 1851.

The position of the $r_{min}$ is defined by F12 as an indicator of the level of segregation of BSSs or dynamical status of the cluster. There is a well established \textit{empirical dynamical clock} relation between $r_{min}$ and core or half-mass relaxation times ($t_{rc}$ or $t_{rh}$) defined by F12. Also, L16 uses $A^{+}_{rh}$ parameter and show that there exists a direct correlation between $r_{min}$ and $A^{+}_{rh}$. Therefore, based on the position of the minima in the BSS radial distribution and $A^{+}_{rh}$ parameter, we suggest that NGC 1851 belongs to \textit{Family II} classification and is an intermediate dynamical state cluster. The bimodal radial distribution suggest that, BSSs located in the outskirts of the cluster are still not affected by the role of dynamical friction. The spatial distribution of the BSSs also suggest that most of the BSSs are located in the central region, while some BSSs are located in the outskirts, which validates our findings. In the outskirts of the GC, NGC 1851, Singh {\em et al.} (2020) found a candidate BSS+EHB binary system, suggesting that some BSSs are located in the outskirts of NGC 1851, where density is low and binary BSSs can form through transfer mass from the companion star.

\section{SUMMARY AND CONCLUSIONS} \label{sec:summary}       
 
\begin{enumerate}

\item We present the dynamical status of the cluster NGC 1851, using data from HST, UVIT/AstoSat, GAIA $DR2$ and ground-based catalogs. All the BSS and reference population used in the study are bonafide members of the cluster.

\item We found in total 172 number of BSSs from the entire cluster sample using NUV-Optical and Optical CMDs for the selection. We also identified 521 and 3324 number of HB and GB stars as  reference populations.

\item The observed radial distribution of BSS shows a peak in the central region, a dip at the intermediate radii ($r_{min}$ $\sim$ $90^{''}$) and a rising trend in the external region. We also estimated $A^{+}_{rh}$ parameter to be $0.391$ $\pm$ $0.006$. The values of both the $r_{min}$ and $A^{+}_{rh}$ parameter therefore indicate that NGC 1851 is an intermediate dynamical state cluster and it belongs to \textit{Family II} classification. This indicates that the BSSs located in the outskirts of the cluster are still not affected by the action of dynamical friction.

\end{enumerate}

\section*{Acknowledgements}
We are thankful to the reviewer for the thoughtful comments and suggestions that improved the quality of the manuscript. This publication uses the data from the AstroSat mission of the Indian Space Research Organisation (ISRO), archived at the Indian Space Science Data Centre (ISSDC). UVIT project is a result of the collaboration between IIA, Bengaluru, IUCAA, Pune, TIFR, Mumbai, several centers of ISRO, and CSA. This publication uses the data from the \textit{ASTROSAT} mission of the Indian Space Research  Organisation  (ISRO),  archived  at  the  Indian  Space  Science  Data Centre (ISSDC). This work has made use of data from the European Space Agency (ESA) mission {\it Gaia} (\url{https://www.cosmos.esa.int/gaia}), processed by the {\it Gaia} Data Processing and Analysis Consortium (DPAC, \url{https://www.cosmos.esa.int/web/gaia/dpac/consortium}). Funding for the DPAC has been provided by national institutions, in particular, the institutions participating in the {\it Gaia} Multilateral Agreement. This research has made use of data, software and/or web tools obtained from the High Energy Astrophysics Science Archive Research Center (HEASARC), a service of the Astrophysics Science Division at NASA/GSFC and of the Smithsonian Astrophysical Observatory's High Energy Astrophysics Division.

\end{document}